\documentclass[a4paper,10pt]{article}
\usepackage{amsmath}
\usepackage{amsfonts}
\usepackage{amssymb}
\usepackage{graphicx}
\begin{document}

\title{  \textsc{  BB mode    spectrum  of  CMB and Inflation}}

\author{N. Malsawmtluangi $ \footnote{e-mail: tei.naulak@uohyd.ac.in}$
        and
        P. K. Suresh $\footnote{e-mail: sureshpk@uohyd.ac.in}$ \\[.2cm]
{\small \it School of Physics, University of Hyderabad. }\\
{\small \it P. O. Central University, Hyderabad 500046. India.}}

\date{\empty}

\maketitle

\begin{abstract}
 Quantum effect on   the BB-mode correlation spectrum of Cosmic Microwave Background for  several inflation models  is studied  with the  BICEP2/Keck Array  and Planck  joint data. The results  do not rule out either  single or multi  field  models of  slow-roll inflation.  The quantum  effect is found  more prominent for the inflation  models with larger values of tensor-to-scalar ratio and smaller values of tensor spectral index.
 \end{abstract}
 
% \pacs{ 04.30.-w, 04.50.-h, 11.25-w, 98.80-k ...............}
%\noindent{\it keywords }: Inflation, gravitational waves,CMB, cosmology

\maketitle

\section{Introduction}
Inflation, a sudden exponential expansion which occurred  for a brief period in the  very early  stage of the universe, is the highly considered  scenario  to overcome  some of the shortcomings of the standard model of cosmology.  There exist  a large number of inflation models by  now  which include single as well as  multi field inflation models \cite{rhb, linde1, linde2, guth1, jm}. All the  models may  agree that  quantum fluctuations during inflation led  to  the  scalar and tensor perturbations. The scalar perturbations represent the density fluctuations which seeded the  galaxy formation in  the universe that we observe today, while the tensor perturbations represent  primordial gravitational waves (GWs). 
The primordial GWs were generated during the inflation by  strong and variable gravitational field of early universe via a mechanism known as  parametric amplification of the zero-point quantum fluctuations which transforms the initial vacuum state with no particle  into a multi-particle quantum state \cite{lpg1, lpg2}, called  squeezed vacuum state \cite{lpg3}. Therefore the  primordial GWs are supposed  to be found  in 
the squeezed vacuum state \cite{lpsq}.
Further, it is believed  that the  tensor fluctuations have  left their own signatures on the cosmic microwave background (CMB) \cite{cmb1, cmb2, cmb3}. Hence  the squeezing  effect  may also be expected to  reflect on the BB-mode correlation  angular power spectrum of CMB \cite{sb}.

Single field slow-roll inflation predicts  almost Gaussian distribution of adiabatic density perturbations with an almost scale invariant spectrum,
whereas  multi field  inflation  generate  non-adiabatic  perturbations giving  rise   to non-Gaussianity.
According to the recent results from Planck mission the "primordial non-Gaussianity is small" \cite{planck1} which tempts to conclude that single field inflation models are favorable over multi field.
However, one of  the attempts \cite{ctb, zak, ale} to explain the hemispherical asymmetry in the CMB sky, hinted by WMAP \cite{eriksen1, hansen, eriksen2} and confirmed by Planck 2013 and 2015 data \cite{planck2, planck3}, 
  indicates that a single field slow roll inflation cannot produce such an asymmetry  without violating 
 the homogeneity of the universe.  
 At same time multi scalar field (most probably inflaton and curvaton) during inflation can produce such an asymmetry  without violating the homogeneity \cite{ale}.  The goal  of the paper is not study the Gaussianity issue, however  the aforementioned  studies show that  it is necessary  to  address  and resolve the  issue of single field or multi field of  inflation  with alternative  approaches   than the Gaussianity or non-Gaussianity test.  Therefore,  the present  work is aimed  to explore  whether the single field and multi field  models of inflation issue can be resolved   by considering the  GWs in the  squeezed vacuum state and then its   effect on  the  $BB$ mode power spectrum of CMB with the BICEP2/Keck Array and Planck collaboration data \cite{bkp}.
 
\section{GW in  squeezed vacuum state}

The perturbed  form of the  Friedmann-Lemaitre-Robertson-Walker (FLRW) metric  for flat  universe is given by
\begin{equation}
ds^2 = S^2(\tau)[-d\tau^2 + (\delta_{ij}+h_{ij})dx^idx^j],
\end{equation}
where $h_{ij}$ is a transverse-traceless perturbation with 
~~~\[\partial_ih^{ij} = 0 , ~~~\delta^{ij}h_{ij}=0,\]
and $|h_{ij}|\ll\delta_{ij}$,where $\delta_{ij}$ is the  space metric, $S$ is the scale factor and $\tau$ is the conformal time. 

The GW  $h_{ij}(\textbf{x},\tau)$ can be decomposed into a collection of Fourier modes as
\begin{eqnarray}\label{fm}
\nonumber h_{ij}(\textbf{x},\tau) = \frac{C}{(2\pi)^\frac{3}{2}} 
 \int_{-\infty}^{+\infty} \frac{d^3\textbf{k}}{\sqrt{2k}}  \sum_{p=1}^2  
 [ h_k^{(p)}(\tau) c_k^{(p)} \varepsilon_{ij}^{(p)}(\textbf{k})e^{i\textbf{k}.\textbf{x}} 
  + h_k^{(p) \ast}(\tau)c_k^{(p) \dagger} \varepsilon_{ij}^{(p) \ast}(\textbf{k})e^{-i\textbf{k}.\textbf{x}} ],
\end{eqnarray}
where $C=\sqrt{16\pi}l_{pl}$  
and  $l_{pl}=\sqrt{G}$ is the Planck length.

The polarization states $\varepsilon_{ij}^{(p)}$, $p=1,2$ are transverse-traceless and symmetric  which obey  the conditions
$\varepsilon_{ij}^{(p)}\delta^{ij}=0$, $\varepsilon_{ij}^{(p)}k^i=0$, $\varepsilon_{ij}^{(p)}\varepsilon^{(p') ij} = 2\delta_{pp'}$, $\varepsilon_{ij}^{(p)}(\textbf{-k})=\varepsilon_{ij}^{(p)}(\textbf{k})$.
These  are linear polarizations, called the plus $(+)$ polarization and cross $(\times)$ polarization. 
The  operators $c_k^{ \dagger}$ and $c_k$ follow  the conditions,
$\left[c_k^{(p)},c_{k'}^{(p') \dagger}\right] = \delta_{pp'}\delta^3(k-k')$,
$\left[c_k^{(p)},c_{k'}^{(p')}\right] = \left[c_k^{(p) \dagger},c_{k'}^{(p') \dagger}\right] = 0$ and 
are governed by the  Heisenberg equations given by
\begin{equation}\label{Hsa}
\frac{d}{d\tau}c_{\textbf{k}}^{\dagger}(\tau) = -i[c_k^{\dagger}(\tau),H],
\end{equation}
\begin{equation}\label{Hsb}
\frac{d}{d\tau}c_{\textbf{k}}(\tau) = -i[c_k(\tau),H].
\end{equation}
The initial vacuum state $|0\rangle$ is defined as
\[c_k^p |0\rangle = 0.\]
The Bogoliubov transformations for the operators $c_{\textbf{k}}^{\dagger}$ and $c_{\textbf{k}}$ are
\begin{equation}
c_{\textbf{k}}^{\dagger}(\tau) = u_k^{\ast}(\tau)c_{\textbf{k}}^{\dagger}(0) + v_k^{\ast}(\tau)c_{\textbf{k}}(0),
\end{equation}
\begin{equation}
c_{\textbf{k}}(\tau) = u_{k}(\tau)c_{\textbf{k}}(0) + v_{k}(\tau)c_{\textbf{k}}^{\dagger}(0),
\end{equation}
where  $c_{\textbf{k}}^{\dagger}(0)$  is the initial value of the creation operator and  $c_{\textbf{k}}(0)$  is   that  for the annihilation operator. The complex functions $u_k(\tau)$ and $v_k(\tau)$  satisfy the following condition
\[|u_k|^2-|v_k|^2=1.\] 
The  mode functions $h_k (\tau)$ with  scale factor $S(\tau)$ can be taken as
\begin{equation}\label{mf}
h_k^{(p)}=\frac{\psi_k^{(p)}}{S}.
\end{equation}
The mode functions can have the following form \footnote{Since the contribution from each polarization is same, here onward, we drop the superscript $(p)$}.
\begin{equation}\label{mff}
\psi_k(\tau) = u_k(\tau) + v_k^{\ast}(\tau),
\end{equation}
which satisfies the equation of  motion
\begin{equation}\label{eom}
\psi_k'' + \left(k^2-\frac{S''}{S}\right)\psi_k = 0.
\end{equation}

Two-point correlation function of the Fourier modes gives the power spectrum of tensor perturbations and can be written as
\begin{equation} \label{tpc}
\langle h_k h^{\ast}_{k'}\rangle = \frac{2\pi^2}{k^3}P_T(k)\delta^3(\textbf{k}-\textbf{k}'),
\end{equation}
where $P_T$ is the gravitational wave power spectrum and $ \langle \, \, \rangle $ denotes ensemble average. 

The gravitational wave  can be written in terms of the mode function and the annihilation and creation operators,  taking the contribution from each polarization to be the same, 
\begin{equation}\label{field}
h(\textbf{x},\tau) = \frac{C}{S(\tau)(2\pi)^\frac{3}{2}}\int_{-\infty}^{+\infty}d^3\textbf{k} [\psi_k(\tau) c_k + \psi_k^{\ast}(\tau)c_k^{\dagger}]e^{i\textbf{k}.\textbf{x}}.
\end{equation}

The squeezed vacuum state can be  defined as \cite{bls}
\begin{equation}
|\xi\rangle = Z(\xi)|0\rangle,
\end{equation}
where the squeezing operator $Z(\xi)$ can be written as 
\begin{equation}
Z(\xi) = \exp\left[\frac{1}{2} \xi^{\ast} b^2 - \frac{1}{2} \xi b^{\dagger 2}\right],
\end{equation}
where $\xi =r_s e^{i \gamma}$, $r_s$ and $\gamma$ are  respectively  the squeezing parameter and  squeezing angle. The action of   the squeezing operator $Z$ on the annihilation and creation operators gives
\begin{eqnarray}\label{sqc}
Z^{\dagger}(\xi)b Z(\xi) &=& b\cosh r_s - b^{\dagger} e^{i\gamma} \sinh r_s, \nonumber \\
Z^{\dagger}(\xi)b^{\dagger} Z(\xi) &=& b^{\dagger}\cosh r_s - b e^{-i\gamma} \sinh r_s. 
\end{eqnarray}

The functions $u_k(\tau)$ and $v_k (\tau)$ in Eq.(\ref{mff}) can be represented in terms of three real functions: the rotation angle $\theta_s$,  the squeezing parameter $r_s$ and  squeezing angle $\gamma$  as
\begin{eqnarray}
u_k &=& e^{i \theta_s} \cosh r_s, \nonumber \\
v_k &=& e^{-i(\theta_s -2\gamma)} \sinh r_s.
\end{eqnarray}

Using Eqns.(\ref{fm}) and (\ref{sqc}), the tensor power spectrum for GWs  in the squeezed vacuum state is obtained  as,
\begin{eqnarray} \label{tpg}
  \langle h_{\bf k}h_{{\bf k}'}^{\ast} \rangle =  \frac{C^2}{S^2} \left[(1+ 2\sinh^2 r_s)|\psi_k|^2 
 +\frac{1}{2} \sinh2r_s (\psi_k^2 e^{i\gamma} + \psi_k^{\ast 2}e^{-i\gamma})\right] 
   \delta^3 (\textbf{k}-\textbf{k}').
\end{eqnarray}
From Eq.(\ref{tpc}) and Eq.(\ref{tpg}), we get the tensor power spectrum in the squeezed vacuum state as
\begin{eqnarray} \label{sqzspc}
P_T(k)= \frac{k^3}{2\pi^2} \frac{C^2}{S^2} \left[(1+ 2\sinh^2 r_s)|\psi_k|^2 + \frac{1}{2} \sinh2r_s (\psi_k^2 e^{i\gamma} + \psi_k^{\ast 2}e^{-i\gamma})\right].
\end{eqnarray}
For  the quasi de Sitter universe, during inflation the conformal time and the scale factor are related as $S(\tau)=\frac{-1}{H\tau (1-\epsilon)}$, where $\epsilon = \frac{m_{pl}^2}{2}\left(\frac{V'}{V}\right)^2$ and $V$ is the potential of the scalar field. For small value of the  slow-roll parameter $\epsilon$, $\vartheta = \frac{3}{2}+\epsilon$, and $n_T = -2\epsilon = 3-2\vartheta$.

For constant $\epsilon$, the equation of motion can  be written as \cite{prl}
\begin{equation} \label{mudd}
\psi''_k + \left[k^2 -\frac{1}{\tau^2}\left(\vartheta^2 - \frac{1}{4}\right)\right]\psi_k =0.
\end{equation}
The general solution for the above equation (Eq.(\ref{mudd})) is
\begin{equation} \label{gsol}
\psi_k(\tau) = \sqrt{-\tau}[C_1 (k)\mathbb{H}^{(1)}_{\vartheta}(-k\tau)+C_2 (k)\mathbb{H}^{(2)}_{\vartheta}(-k\tau)],
\end{equation}
where $\mathbb{H}^{(1)}_{\vartheta}$ and $\mathbb{H}^{(2)}_{\vartheta}$ are the Hankel functions of the first and second kind, and $C_1$ and $C_2$ are the integration constants.

Within the horizon ($k>>SH$), the modes can be approximated using the flat spacetime solutions as \[\psi^0_k(\tau)=\frac{1}{\sqrt{2k}}e^{-ik\tau}.\]

Using the above solution, the constants of integration become
\begin{eqnarray}
C_1(k)&=&\frac{\sqrt{\pi}}{2}\exp\left[i\left(\vartheta+\frac{1}{2}\right)\left(\frac{\pi}{2}\right)\right], \nonumber \\ 
C_2(k)&=&0.
\end{eqnarray}
Hence  for long wavelength limiting case  ($k<<S H$), Eq.(\ref{gsol}) gives
\begin{equation} \label{musol}
\psi_k(\tau)=e^{i\left(\vartheta-\frac{1}{2}\right)\left(\frac{\pi}{2}\right)}2^{\vartheta-\frac{3}{2}}\frac{\Gamma(\vartheta)}{\Gamma{\left(\frac{3}{2})\right)}}\frac{1}{\sqrt{2k}}(-k\tau)^{\frac{1}{2}-\vartheta}.
\end{equation}
Using Eq.(\ref{musol}) in Eq.(\ref{sqzspc}), the tensor  power spectrum for GWs in the superhorizon limit ($k<<S H$) is,
\begin{eqnarray}
\nonumber  P_T(k) = C^2\left(\frac{H}{2\pi}\right)^2  \left(\frac{k}{S H}\right)^{3-2\vartheta} 
  \left[1+2\sinh^2 r_s   + \sinh2r_s \cos\left(\gamma + \left(\vartheta-\frac{1}{2}\right)\pi\right)\right].
\end{eqnarray}

By taking $A_T(k_0) = C^2 \left(\frac{H_{k0}}{2\pi}\right)^2$ with  $H_{k0}$, the Hubble parameter, at $SH=k_0$ during the inflation, $k_0$ is  the pivot wavenumber,   the  tensor power spectrum  in terms of the tensor spectral index $n_T$  is obtained as
\begin{eqnarray} \label{gwsvs}
 \nonumber P_T(k) = A_T(k_0) \left(\frac{k}{k_0}\right)^{n_T}  
  \left[1+2\sinh^2 r_s  +  \sinh2r_s \cos\left(\gamma + (2-n_T)\frac{\pi}{2}\right)\right],
\end{eqnarray}
which  is  the   tensor power spectrum  of GWs in the squeezed vacuum state.

All GW modes start in the same vacuum state with $r_s = 0$ initially.  However,  as mentioned earlier, quantum mechanical  evolution under parametric influence transforms the initial vacuum state into  strongly squeezed  vacuum state \cite{lpgx}. The variance of the mode's phase is strongly squeezed whereas the variance of its amplitude is being enhanced so that the uncertainty product remains intact. The parameter of squeezing grows all the way up in the amplifying regime and can vary from zero in the vacuum state up to a very large value by the end of the amplifying regime. Its value for the present epoch is  $1.2 \times 10^{-2}$ \cite{lpg5}. For the inflationary period, it is  up to 1 ($0 \le r_s \le 1$) which, we use  for the present study.

\section{Slow-roll inflationary scenario}
In the simplest inflationary scenario,  a homogeneous scalar field known as inflaton  $\phi$,  drives the accelerated expansion of early universe.

The equation of motion for the  inflaton can be written as
\begin{equation}
\ddot{\phi} + 3H \dot{\phi} + V' (\phi)=0,
\end{equation}
where the Hubble parameter  is determined by the energy density of the field,\[ \rho_{\phi} = \frac{\dot{\phi}^2}{2}+V,\] so that the Friedmann equation becomes
\begin{equation}\label{fdmeq}
H^2 = \frac{1}{3 m^2_{pl}}\left( \frac{1}{2} \dot{\phi}^2 + V\right).
\end{equation}
Under  the slow-roll limiting condition the Friedmann equation takes the following form 
\begin{equation}
H^2 \simeq \frac{V}{3m^2_{pl}}.
\end{equation}
 This can be  characterized  with  the slow-roll parameters which are defined  as 
\begin{eqnarray}
\epsilon &\equiv & \frac{m_{pl}^2}{2}\left(\frac{V'}{V}\right)^2, \nonumber \\ \eta &\equiv & m^2_{pl}\left(\frac{V''}{V}\right),
\end{eqnarray}
and so on. Inflation lasts as long as  $\epsilon \ll 1$ and $|\eta|\ll 1$ and it keeps the Hubble rate nearly constant. The slow-roll parameters  can be used to study the fluctuations generated during inflation.

The parameter that measures the relative strength of the tensor spectrum to the scalar spectrum is given by
\begin{equation}\label{r}
r \equiv \frac{P_T (k)}{P_S (k)} \simeq 16\epsilon,
\end{equation}
which is known as the tensor-to-scalar ratio
and is very useful to distinguish different models of   inflation.  For the present study the scalar power spectrum is taken to be $P_S= 2.43 \times 10^{-9}$.

\section{Slow-roll inflation models}
In this section, we discuss several slow-roll inflation models  briefly for which 
the upper bound for the tensor-to-scalar ratio is  $r<0.07$ and lower limit is $r \simeq \mathcal{O}(10^{-3})$ and  are constrained with various  CMB observations \cite{gb}. We compute the slow roll parameters, tensor spectral index and  tensor power spectrum  for the following slow-roll inflation models.

\subsection{Natural inflation model}
In this model the inflaton is considered to be the pseudo-Nambu Goldstone boson which arises whenever global symmetry is broken \cite{ni1, ni2}.

The corresponding potential of the inflaton is given by
\begin{equation}
V(\phi) = M^4 \left[1 + \cos\left(\frac{\phi}{f}\right)\right],
\end{equation}
where $f/m_{pl} = 10^2$ is the energy scale at which symmetry is broken, $M/m_{pl} = 10^{-2}$.

The obtained   parameters  for the Natural inflation model are
\begin{eqnarray*}
r &=& 2.06 \times 10^{-2}, \\
\epsilon &=& 1.29 \times 10^{-3},\\
\eta &=& 1.24 \times 10^{-3},\\
n_T &=& -2.58 \times 10^{-3}.
\end{eqnarray*}
The  initial tensor power spectrum for the model is obtained as $P_T = 5.027 \times 10^{-11}$.

\subsection{Arctan inflation model}
This model is often studied as a toy model \cite{ai1, ai2}. In the present study, we consider a large field model which starts at a large value and then evolves to a minimum at the origin.

The potential for the model is
\begin{equation}
V(\phi) = M^4 \left[1 - \arctan\left(\frac{\phi}{\mu}\right)\right],
\end{equation}
where $\mu/m_{pl} = 10^{-2}$ is a free parameter which characterizes the typical vacuum expectation value at which inflation takes place, $M/m_{pl} = 10^{-3}$.

 Thus the parameters for the Arctan inflation model are obtained as
\begin{eqnarray*}
r &=& 1.38 \times 10^{-2}, \\
\epsilon &=& 8.62 \times 10^{-4},\\
\eta &=& 3.0 \times 10^{-2},\\
n_T &=& -1.72 \times 10^{-3}.
\end{eqnarray*}
The initial tensor power spectrum for the model is obtained as, $P_T = 3.35 \times 10^{-11}$.

\subsection{Inverse monomial inflation model}
This model is studied  in the context of quintessential inflation \cite{im1, im2, im3}; the inflaton need not necessarily decay and reheating arises naturally even when the potential does not have a global minimum, radiation is created via gravitational particle production.

The inflaton potential for the model is
\begin{equation}
V(\phi) = M^4 \left(\frac{\phi}{m_{pl}}\right)^{-p},
\end{equation}
where $p$ is a positive parameter, $M/m_{pl} = 10^{-1}$.

The corresponding calculated parameters are:
\begin{eqnarray*}
r &=& 2.0 \times 10^{-3}, \\
\epsilon &=& 1.25 \times 10^{-4},\\
\eta &=& 3.33 \times 10^{-4},\\
n_T &=& -2.50 \times 10^{-4}.
\end{eqnarray*}
The corresponding initial tensor power spectrum for this model is, $P_T = 4.86 \times 10^{-12}$.

\subsection{Loop inflation model}
In this scenario, the flatness of the inflaton potential is altered by symmetry breaking which produces quantum radiative corrections in which one loop order correction takes the form of a logarithmic function \cite{li1, li2, li3}.

The potential for the loop inflation model can be written as
\begin{equation}
V(\phi) = M^4 \left[1 + \alpha \ln \left(\frac{\phi}{m_{pl}}\right)\right],
\end{equation}
where $\alpha = g^2/16\pi^2$ tunes the strength of radiative effects, $M = 10^{16}$ GeV.

The corresponding  calculated parameters are
\begin{eqnarray*}
r &=& 4.34 \times 10^{-2}, \\
\epsilon &=& 3.09 \times 10^{-3},\\
\eta &=& -2.06 \times 10^{-2},\\
n_T &=& -6.18 \times 10^{-3}.
\end{eqnarray*}
The initial tensor power spectrum is obtained as, $P_T = 1.2 \times 10^{-10}$.

\subsection{Coleman-Weinberg inflation model}
The potential in this scenario is introduced in the context of spontaneous symmetry breaking generated by radiative corrections \cite{cw1, cw2, cw3}.

The potential for the model is 
\begin{equation}
V(\phi) = M^4 \left[1 + \alpha \left(\frac{\phi}{\sigma}\right)^4 \ln\left(\frac{\phi}{\sigma}\right)\right],
\end{equation}
where $\alpha = 4e$, $M = 10^{16}$ GeV, $\sigma = 10 m_{pl}$ sets the typical vacuum expectation value at which inflation takes place.

The parameters are obtained as
\begin{eqnarray*}
r &=& 7.77 \times 10^{-3}, \\
\epsilon &=& 4.86 \times 10^{-4},\\
\eta &=& -4.42 \times 10^{-2},\\
n_T &=& -9.72 \times 10^{-4}.
\end{eqnarray*}
The corresponding initial tensor power spectrum is, $P_T = 1.89 \times 10^{-11}$.

\subsection{Quadratic chaotic inflation model with radiative corrections}
This model is a simple quadratic chaotic inflation model \cite{ch1} studied under the assumption that the scalar field interacts with the fermion field thus leading to the quantum radiative correction which takes the form of a logarithmic function \cite{ch2}.

The potential with radiative correction  is
\begin{equation}
V(\phi) = \frac{1}{2}m^2\phi^2 - \frac{g^4}{16\pi^2}\phi^4\ln\left(\frac{\phi}{m_{pl}}\right),
\end{equation}
where $g$ is the Yukawa coupling and  $m = 3.44 \times 10^{12}$ GeV.

The associated parameters are obtained as
\begin{eqnarray*}
r &=& 2.98 \times 10^{-2}, \\
\epsilon &=& 1.86 \times 10^{-3},\\
\eta &=& 1.86 \times 10^{-3},\\
n_T &=& -3.72 \times 10^{-3}.
\end{eqnarray*}
The initial tensor power spectrum for this model is obtained as, $P_T = 7.25 \times 10^{-11}$.

\subsection{Hybrid inflation model}
This is a multi-scalar field \cite{hy1, hy2} model where $\phi$ field drives the inflation and the symmetry breaking of $\sigma$ field triggers the end of inflation.

The potential of  hybrid inflation model is 
\begin{equation}
V = \frac{1}{4\lambda}(M^2-\lambda \sigma^2)^2 + \frac{1}{2}m^2\phi^2 + \frac{1}{2}g^2\phi^2\sigma^2,
\end{equation}
where $M = 1.21 \times 10^{16}$ GeV, $m = 3.65 \times 10^{11}$ GeV, $\lambda = 1$, $g = 8 \times 10^{-4}$.

The various obtained parameters  of the model are
\begin{eqnarray*}
r &=& 4.24 \times 10^{-3}, \\
\epsilon &=& 2.65 \times 10^{-4},\\
\eta &=& 1.47 \times 10^{-4},\\
n_T &=& -5.3 \times 10^{-4}.
\end{eqnarray*}
The initial tensor power spectrum for the model is obtained as, $P_T = 1.03 \times 10^{-11}$.

\section{BB-mode   correlation angular power spectrum}
The  $BB$-mode correlation  angular power spectrum of  CMB  in terms of the multipole moments $l$ is given by \cite{cmb4, cmb5}
\begin{eqnarray}
  \nonumber C_l^{BB} = (4\pi)^2 \int dk k^2 P_T(k) 
  \left| \int_0^{\tau_0} d\tau g(\tau) h_k(\tau) 
   \Big\{(8x + 2x^2 \partial_x)\frac{j_l(x)}{x^2}\Big\}_{x=k(\tau_0-\tau)}\right|^2
\end{eqnarray}
where $g(\tau) = \dot{\kappa} \, e^{-\kappa}$ is the probability distribution of the last scattering with the  optical depth  $\kappa$   and $j_l(x)$ is the spherical Bessel function.

The $BB$-mode correlation angular  power spectrum of  CMB   for the aforemetioned slow-roll inflationary models is obtained. The angular  power spectrum of CMB  for the  inflation models  is generated using the CAMB code with  $n_T$ corresponding to each model.  For all inflation models, the optical depth is taken as $\kappa = 0.08$, the pivot wave number for tensor mode   and scalar mode respectively  taken as $k_0 =0.002$ Mpc$^{-1}$ and  $k_0 =0.05$ Mpc$^{-1} $. 
The obtained results are  compared with the  joint analysis data  of BICEP2/Keck Array and Planck. The limit (BK x BK - $\alpha$  BK x P)/(1 - $\alpha$) at $\alpha = \alpha_{fid} = 0.04$    is taken after the subtraction of the dust contribution (which is 0.04 times as much in the BICEP2 band as it is in the Planck 353 GHz band).

\begin{figure}
\includegraphics[scale=0.33]{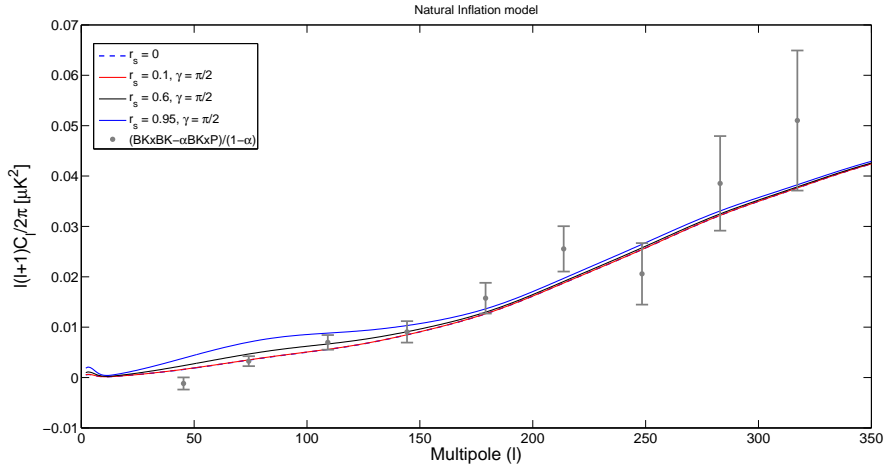}
\caption{$BB$-mode correlation angular power spectrum of CMB  for the Natural inflation model for various values of squeezing parameter with joint data of  BICEP2/ Keck Array  and Planck.}\label{f1}
\end{figure}

\begin{figure}
\includegraphics[scale=0.33]{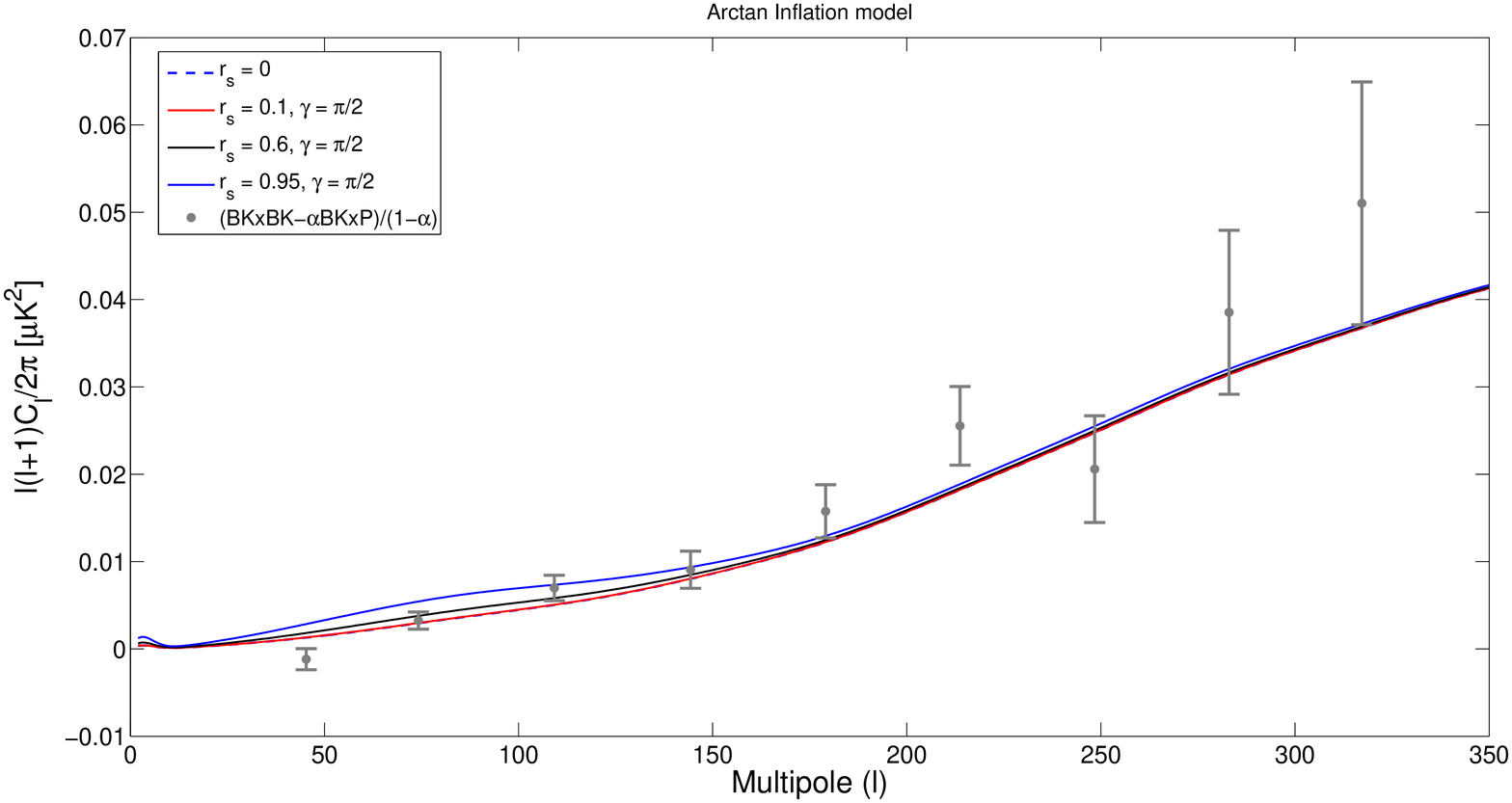}
\caption{$BB$-mode correlation angular power spectrum of CMB  for  the Arctan inflation model for various values of squeezing parameter with joint data of  BICEP2/ Keck Array  and Planck.}\label{f2}
\end{figure}

\begin{figure}
\includegraphics[scale=0.33]{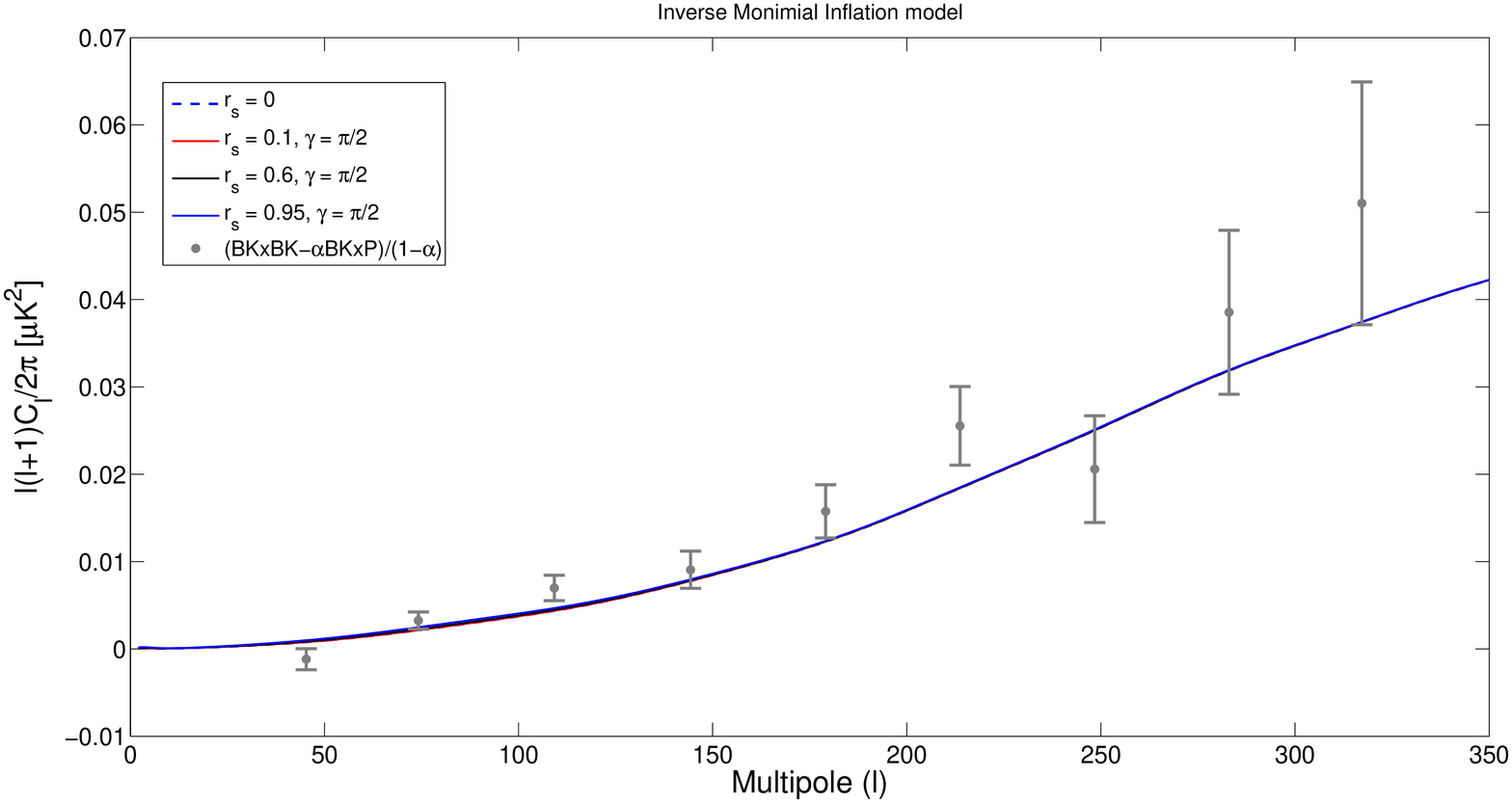}
\caption{$BB$-mode correlation angular power spectrum of CMB   for  the Inverse monomial inflation model for various values of squeezing parameter with joint data of  BICEP2/ Keck Array  and Planck.}\label{f3}
\end{figure}

\begin{figure}
\includegraphics[scale=0.33]{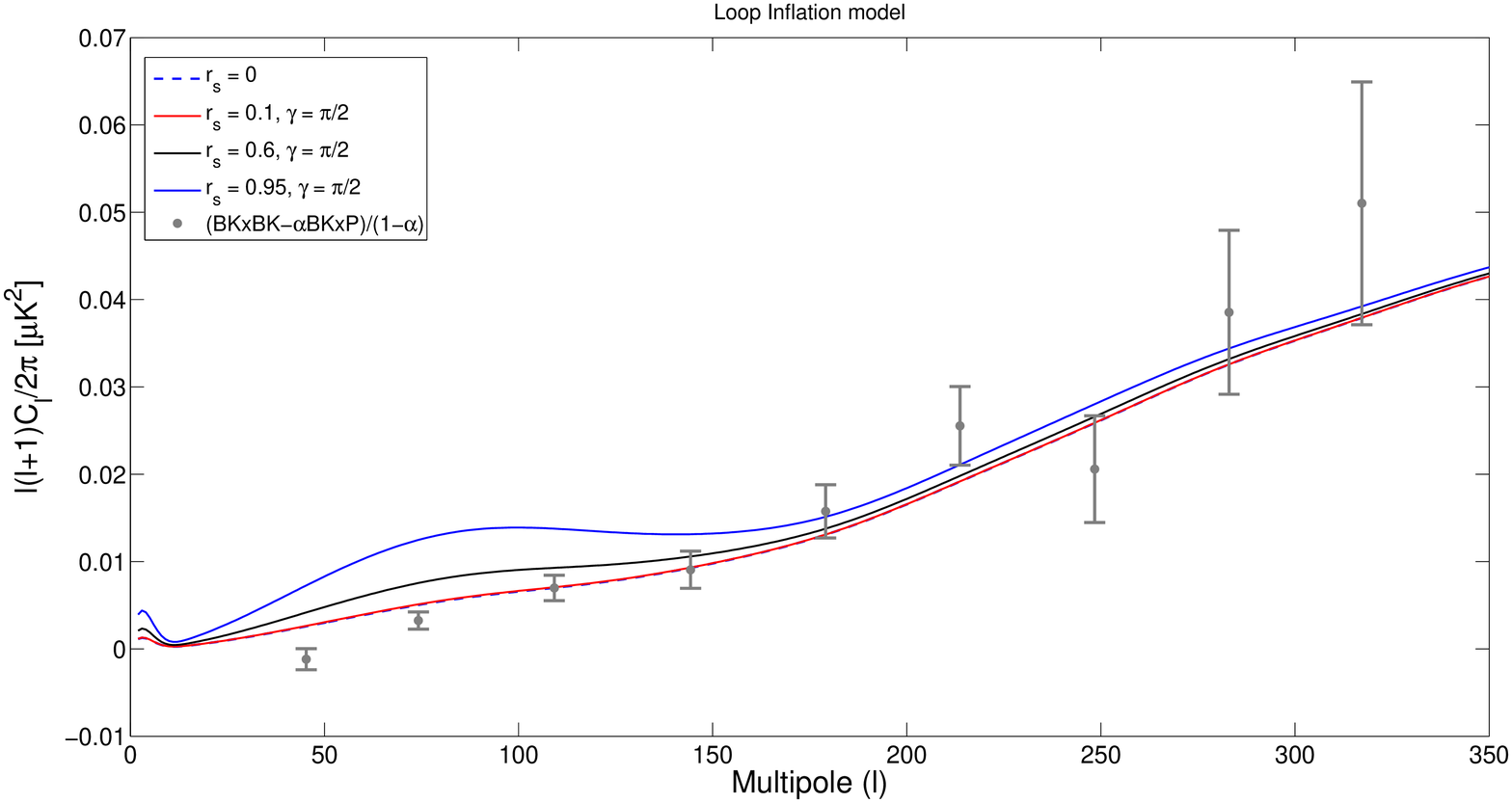}
\caption{$BB$-mode correlation angular power spectrum of CMB   for the Loop inflation model for various values of squeezing parameter with joint data of  BICEP2/ Keck Array  and Planck.}\label{f5}
\end{figure}

\begin{figure}
\includegraphics[scale=0.33]{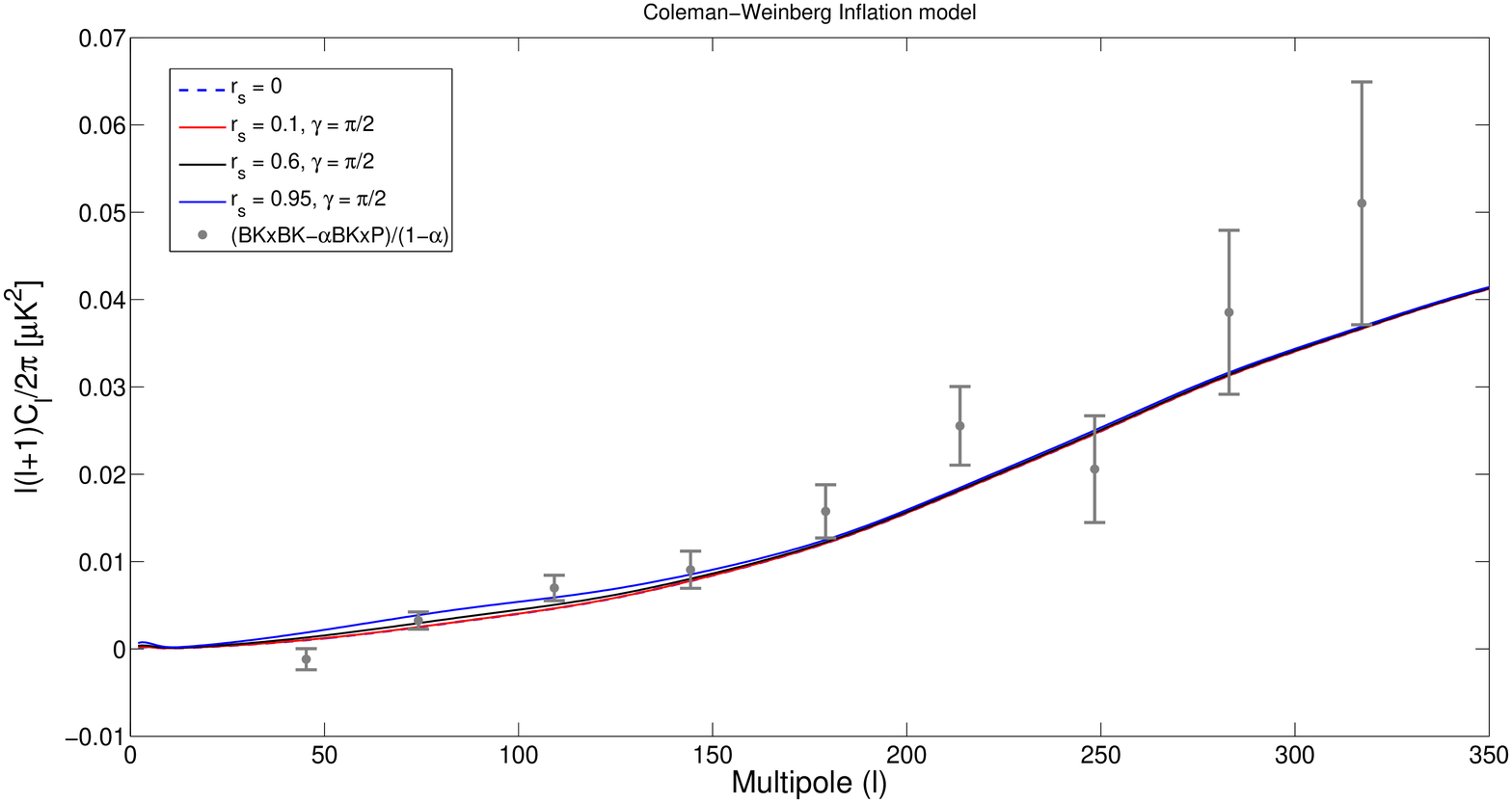}
\caption{$BB$-mode correlation angular power spectrum of CMB   for the Coleman-Weinberg inflation model for various values of squeezing parameter with joint data of  BICEP2/ Keck Array  and Planck.}\label{f6}
\end{figure}

\begin{figure}
\includegraphics[scale=0.33]{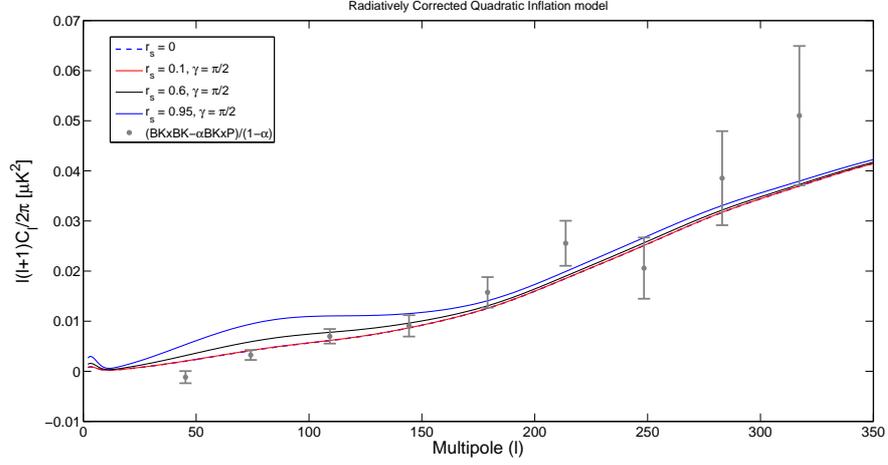}
\caption{$BB$-mode correlation angular power spectrum of CMB   for the Quadratic chaotic inflation model with radiative corrections for various values of squeezing parameter with joint data of  BICEP2/ Keck Array  and Planck.}\label{f7}
\end{figure}

\begin{figure}
\includegraphics[scale=0.33]{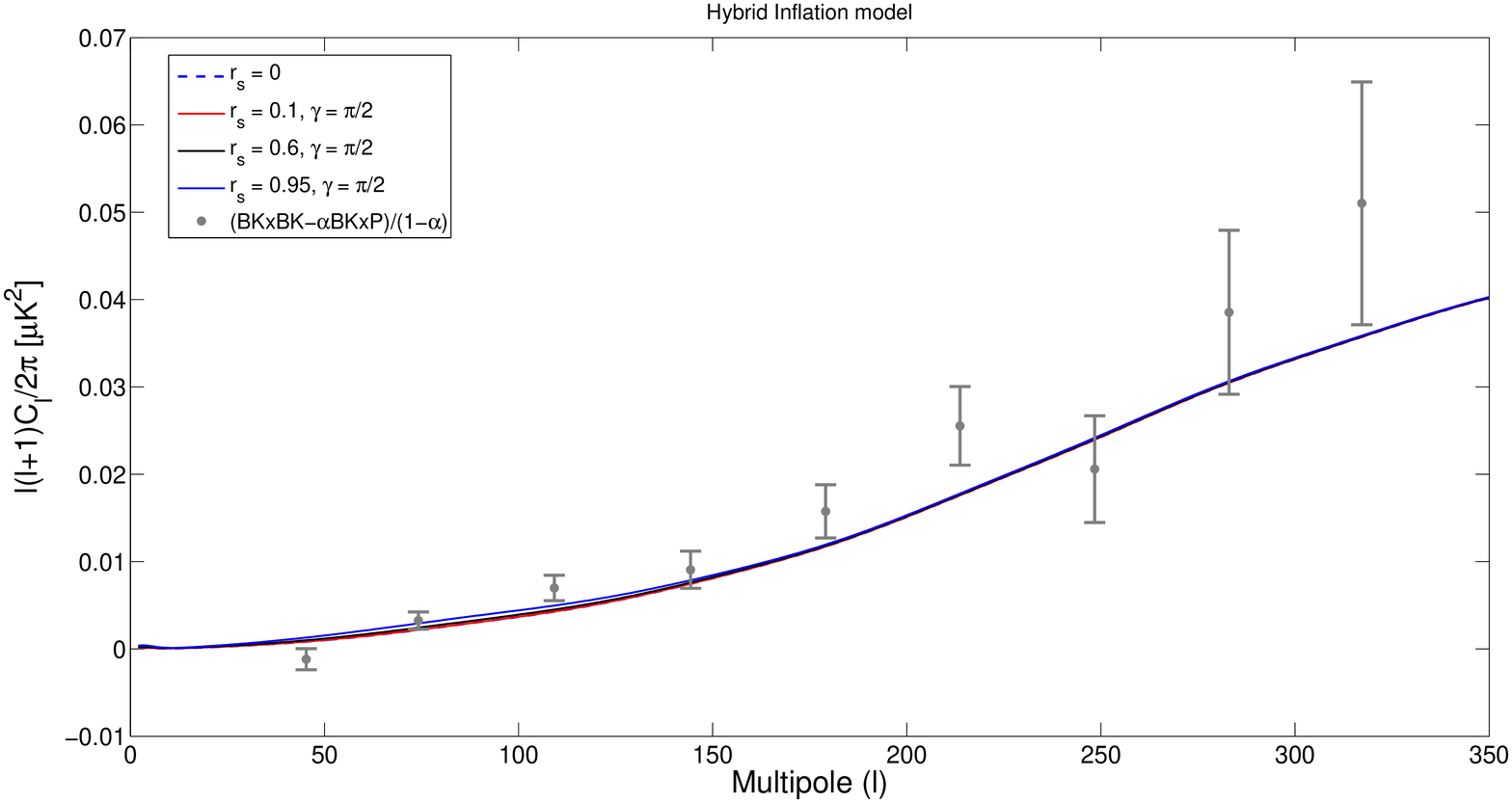}
\caption{$BB$-mode correlation angular power spectrum of CMB   for the Hybrid inflation model for various values of squeezing parameter with joint data of  BICEP2/ Keck Array  and Planck.}\label{f4}
\end{figure}
The obtained $BB$-mode correlation angular power spectrum of CMB  for the different inflation models  with   various  values of squeezing parameter  are studied  with the  BICEPT2/Keck Array and Planck joint data  and results  are given in Figs.\ref{f1}, \ref{f2}, \ref{f3}, \ref{f5}, \ref{f6}, \ref{f7} and \ref{f4}.

\section{Discussion and conclusion}
The $BB$ mode correlation angular power spectrum of CMB  for several slow-roll  inflation models  is studied by considering the primordial gravitational in the squeezed vacuum state.  The  obtained $BB$ mode correlation angular power spectra for different inflation models  are found  within  the constraint of  recent collaboration data of BICEP2/ Keck Array at 150 GHz and Planck 353 GHz. 
Note that higher multipoles (smaller angles) are ignored for the analysis due to contamination from lensing effect.  It can be  observed that   the $BB$ mode angular spectrum gets enhanced with increase in squeezing parameter value  for all the  inflation models. Further, larger the deviation from scale invariance, stronger is the squeezing effect. That is, for models with larger values of tensor-to-scalar ratio and smaller values of tensor spectral index,  the squeezing effect is found  more prominent.  These  show the role of quantum phenomena  on the primordial GWs and consequently on the  BB mode power spectrum of CMB.

 Note that, all the inflationary models that  are considered  in the present work  are  large single field models except the  Hybrid inflation model which is a multi-field  model. 
The recent  results from Planck mission  
 favour single field inflation models.  The attempts  that are made 
 to explain the observed  hemispherical symmetry in the CMB sky  with   single field slow roll inflation  turns out  to be unsuccessful because it  cannot produce such an asymmetry  without violating the homogeneity of the universe but  multi field can generate such asymmetry. The present work made an attempt   
to resolve the tension between  single field and multi field  models of inflation  based on  the $BB$ mode spectrum of CMB. However, the results of  the present study on $BB$ mode correlation angular power spectrum of CMB for various slow-roll models of inflation  with the current joint  data of  BICEP2/ Keck Array  and Planck also  do not rule out either  single or multi field scalar field  models of inflation. From these studies it may be concluded that   studies at very fundamental  level  is required to  understand the multi or single filed issue of  inflation models.   Various cosmological observations and further study may resolve   this important issues in nearby future.

\section*{Acknowledgments}
N M acknowledges the financial support of Council of Scientific and Industrial Research (CSIR), New Delhi. P K S thanks  SERB, New Delhi for the financial support. The authors would like to thank  Planck and BICEP2 websites for the data.

 %\section*{References}

\end{document}